\documentclass[12pt]{article}
\usepackage{indentfirst}
\usepackage{CJK}
\usepackage{graphicx}

\begin{document}

\title{Isospin splitting of nucleon effective mass and symmetry energy in isotopic nuclear reactions}

\author{Ya-Fei Guo$^{1,2}$, Peng-Hui Chen$^{1,2}$, Fei Niu$^{1,3}$, Hong-Fei Zhang$^{2}$,   \\
Gen-Ming Jin$^{1}$, Zhao-Qing Feng$^{1}$
\footnote{Corresponding author. Tel. +86 931 4969152.  \newline \emph{E-mail address:} fengzhq@impcas.ac.cn (Z.-Q. Feng)}}
\date{}
\maketitle

\begin{center}
\small \emph{$^{1}$Institute of Modern Physics, Chinese Academy of Sciences, Lanzhou 730000, People's Republic of China   \\
$^{2}$School of Nuclear Science and Technology, Lanzhou University, Lanzhou 730000, People¡¯s Republic of China   \\
$^{3}$Institute of Particle and Nuclear Physics, Henan Normal University, Xinxiang 453007, People's Republic of China}
\end{center}

\textbf{Abstract}
\par
Within an isospin and momentum dependent transport model, the dynamics of isospin particles (nucleons and light clusters) in Fermi-energy heavy-ion collisions are investigated for constraining the isospin splitting of nucleon effective mass and the symmetry energy at subsaturation densities. The mass splitting of $m^{*}_{n}>m^{*}_{p}$ and $m^{*}_{n}<m^{*}_{p}$ in nuclear matter and the different stiffness of symmetry energy are used in the model. The single and double neutron to proton ratios of free nucleons and light particles are thoroughly investigated in the isotopic nuclear reactions of $^{112}$Sn+$^{112}$Sn and $^{124}$Sn+$^{124}$Sn at the incident energies of 50 and 120 MeV/nucleon, respectively. It is found that the both effective mass splitting and symmetry energy impact the kinetic energy spectra of the single ratios, in particular at the high energy tail (larger than 20 MeV). Specific constraints are obtained from the double ratio spectra, which are evaluated from the ratios of isospin observables produced in $^{124}$Sn+$^{124}$Sn over $^{112}$Sn+$^{112}$Sn collisions. A mass splitting of $m^{*}_{n}<m^{*}_{p}$ is constrained from the available data at the energy of 120 MeV/nucleon. A soft symmetry energy with the stiffness of $\gamma_{s}=$0.5 is close to the experimental double ratio spectra at both energies.


\bigskip

The mass of a nucleon in nuclear matter is different with the in-vacuum case due to the interaction with its surrounding nucleons \cite{Ma01,Da02}. In the neutron-rich nuclear matter, there exists a splitting of neutron and proton effective masses (the nonrelativistic mass or Landau mass). The strength increases with the isospin asymmetry and the nucleon density. There are wide difference for the predictions of the isospin splitting of nucleon effective mass based on nuclear many-body theories. For example, calculations using Landau-Fermi-liquid theory \cite{Sj03} and the nonrelativistic Brueckner-Hartree-Fock theory \cite{Zu04} present a neutron-proton mass splitting of $m^{*}_{n}>m^{*}_{p}$, whereas other calculations using relativistic mean-field theory and relativistic Dirac-Brueckener theory \cite{Li05,Ho06} give a contrary result. The Skyrme-Hartree-Fock approach predicts both splitting of $m^{*}_{n}>m^{*}_{p}$ and $m^{*}_{n}<m^{*}_{p}$ exist with different Skyrme parameters \cite{Ba07}.
The nucleon Landau mass splitting in neutron-rich matter results from the momentum-dependence of the symmetry potential, which directly affects the isospin transport in heavy-ion collisions and consequently the extraction of the density dependence of the symmetry energy. Some observables were proposed for extracting the isospin splitting of nucleon effective mass, i.e., the neutron/proton ratio at high momenta or kinetic energies, elliptic flows in the domain of mid-rapidities, elliptic flow difference between neutrons and protons at high momentum tail etc \cite{Gi10,Fe12}.

There are a lot of works on the constraints of the density dependence of the symmetry energy from heavy-ion collisions, which has important application in nuclear physics \cite{Ba07,Gi10,Li08,Fe16,Fa06} and also in astrophysics \cite{Po11,Ba12}. It is possible to use the heavy-ion reactions with neutron-rich beams to explore the density-dependent symmetry energy and the effective mass splitting of neutron and proton in the nuclear matter. In this work, we investigate the effective mass splitting of neutron and proton and the density-dependent symmetry energy at sub-saturation densities in heavy-ion collisions within the Lanzhou quantum molecular dynamics (LQMD) transport model. The momentum dependence of the symmetry potential was included in the model, which results in the splitting of the nucleon effective mass in nuclear medium \cite{Fe12b}. The isospin dynamics of fast nucleons and light clusters is concentrated on.

In the LQMD model, the temporal evolutions of the baryons (nucleons and resonances) and mesons in the reaction system under the self-consistently generated mean-field are governed by Hamilton's equations of motion. Based on the Skyrme interactions, we constructed an isospin-, density-, and momentum-dependent potentials originated from the Hamiltonian, which consists of the relativistic energy, the effective interaction and the momentum-dependent potentials. The effective interaction potential is composed of the Coulomb potential and the local interaction.

The local interaction potential is derived from the energy-density functional as the form of
$U_{loc}=\int V_{loc}(\rho(\mathbf{r}))d\mathbf{r}$. The functional reads
\begin{eqnarray}
V_{loc}(\rho)=&& \frac{\alpha}{2}\frac{\rho^{2}}{\rho_{0}} +
\frac{\beta}{1+\gamma}\frac{\rho^{1+\gamma}}{\rho_{0}^{\gamma}} + E_{sym}^{loc}(\rho)\rho\delta^{2}
\nonumber \\
&& + \frac{g_{sur}}{2\rho_{0}}(\nabla\rho)^{2} + \frac{g_{sur}^{iso}}{2\rho_{0}}[\nabla(\rho_{n}-\rho_{p})]^{2},
\end{eqnarray}
where the $\rho_{n}$, $\rho_{p}$ and $\rho=\rho_{n}+\rho_{p}$ are the neutron, proton and total densities, respectively, and the $\delta=(\rho_{n}-\rho_{p})/(\rho_{n}+\rho_{p})$ being the isospin asymmetry. The parameters $\alpha$, $\beta$, $\gamma$, $g_{sur}$ $g_{sur}^{iso}$ and $\rho_{0}$ are taken to be -215.7 MeV, 142.4 MeV, 1.322, 23 MeV fm$^{2}$, -2.7 MeV fm$^{2}$ and 0.16 fm$^{-3}$, respectively. A compression modulus of K=230 MeV for isospin symmetric nuclear matter is obtained with these parameters. The local part $E_{sym}^{loc}(\rho)=\frac{1}{2}C_{sym}(\rho/\rho_{0})^{\gamma_{s}}$ with $\gamma_{s}$=0.5, 1 and 2 lead to the soft, linear and hard symmetry energies, respectively. The parameter $C_{sym}$ is taken as the values of 52.5 MeV and 23.5 MeV for the effective mass splittings of $m^{*}_{n}>m^{*}_{p}$ and $m^{*}_{n}<m^{*}_{p}$, respectively, which leads to the symmetry energy of 31.5 MeV at saturation density.

The nucleon effective mass in nuclear medium is contributed from the momentum-dependent interaction in the LQMD model. A Skyrme-type momentum-dependent potential is used in the LQMD model \cite{Fe11}
\begin{eqnarray}
U_{mom}=&& \frac{1}{2\rho_{0}}\sum_{i,j,j\neq i}\sum_{\tau,\tau'}C_{\tau,\tau'}\delta_{\tau,\tau_{i}}\delta_{\tau',\tau_j}\int\int\int d\textbf{p}d\textbf{p}'d\textbf{r} \nonumber \\
&& \times f_{i}(\textbf{r},\textbf{p},t)\left[\texttt{ln}(\epsilon(\textbf{p}-\textbf{p}')^{2}+1)\right]^{2}f_{j}(\textbf{r},\textbf{p}',t).
\end{eqnarray}
Here $C_{\tau,\tau}=C_{mom}(1+x)$, $C_{\tau,\tau'}=C_{mom}(1-x)$ and the isospin symbols $\tau(\tau')$ represent proton or neutron. The parameters $C_{mom}$ and $\epsilon$ was determined by fitting the real part of optical potential as a function of incident energy from the proton-nucleus elastic scattering data. The effective (Landau) mass in nuclear matter is calculated through the potential as $m_{\tau}^{\ast}=m_{\tau}/ \left(1+\frac{m_{\tau}}{|\textbf{p}|}|\frac{dU_{\tau}}{d\textbf{p}}|\right)$ with the in-vacuum mass $m_{\tau}$ at Fermi momentum of $\textbf{p}=\textbf{p}_{F}$. Therefore, the nucleon effective mass only depends on the momentum-dependent interactions. In the calculation, we take the values of 1.76 MeV, 500 c$^{2}$/GeV$^{2}$ for the $C_{mom}$ and $\epsilon$, respectively, which result in the effective mass $m^{*}/m$=0.75 in nuclear medium at saturation density for symmetric nuclear matter. The parameter $x$ is changed as the strength of the mass splitting, and the values of -0.65 and 0.65 are respective to the cases of $m^{*}_{n}>m^{*}_{p}$ and $m^{*}_{n}<m^{*}_{p}$, respectively.

%%%%%%%%%%%%%%%%%%%%%%%%%%%%%%%%%%%%%%% figure 1 %%%%%%%%%%%%%%%%%%%%%%%%
\begin{figure}
\begin{center}
{\includegraphics*[width=0.8\textwidth]{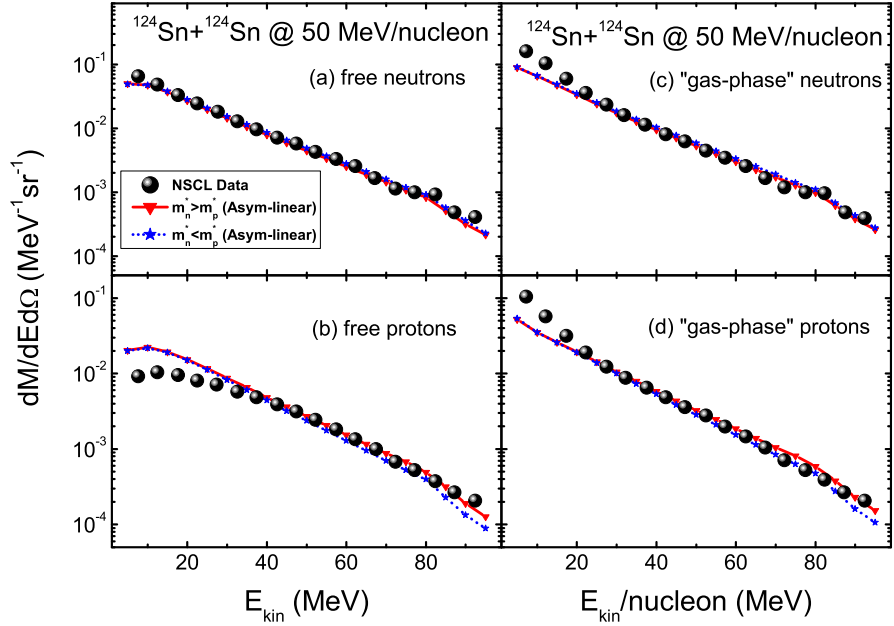}}
\end{center}
\caption{ Kinetic energy spectra of free nucleons [panels (a) and (b)], gas-phase nucleons (nucleons, hydrogen and helium isotopes) [panels (c) and (d)] in the $^{124}$Sn+$^{124}$Sn reactions at an incident energy of 50 MeV/nucleon in the impact parameter domain of 0-3 fm. The experimental data from NSCL \cite{Co16} are shown for comparison.}
\end{figure}
%%%%%%%%%%%%%%%%%%%%%%%%%%%%%%%%%%%%%%%%%%%%%%%%%%%%%%%%%%%%%%%%%%%

The dynamics of isospin particles (nucleons and light clusters) produced in heavy-ion collisions is influenced by the isospin dependent interactions in the mean-field potentials, which could be used as observables for extracting the density dependence of symmetry energy. The Fermi-energy heavy-ion collisions have the longer isospin relaxation time, which lead to the pronounced isospin effect in comparison to the high-density probes, such as the $\pi^{-}/\pi^{+}$, $K^{0}/K^{+}$, $\Sigma^{-}/\Sigma^{+}$ etc \cite{Fe10,Fe13}. It has been found that the
momentum-dependent potential plays an important role on the fast nucleon emissions in heavy-ion collisions \cite{Fe12,Fe11}. The preequilibrium neutron and proton transverse emission in isotopic reaction systems at an incident energy of 50 MeV/A was measured at the National Superconducting Cyclotron Laboratory
(Michigan State University, East Lansing, MI, USA) \cite{Fa06}. Recently, the data were updated with less statistical errors \cite{Co16}. Shown in Fig. 1 is the kinetic energy spectra of free nucleons and gas-phase nucleons (nucleons, hydrogen and helium isotopes) in collisions of $^{124}$Sn+$^{124}$Sn at a beam energy of 50 MeV/nucleon with the effective mass splittings of $m^{*}_{n}>m^{*}_{p}$ and $m^{*}_{n}<m^{*}_{p}$, respectively. The particles emitted from the impact parameter domain of 0-3 fm and perpendicular to the beam direction with the polar angle cut of $70^{o}<\theta<110^{o}$ ($\cos\theta=p_{z}/\sqrt{p_{x}^{2}+p_{y}^{2}+p_{z}^{2}}$) are analyzed. We count the gas-phase proton and neutron yields by multiplying the proton and neutron numbers on the fragments, respectively. It is interest found that the neutron spectra is not sensitive to the isospin splitting of nucleon effective mass. However, the proton yields depend on the splitting, in particular at the high kinetic energies. It is caused from the fact that the momentum-dependent interaction of $m^{*}_{n}>m^{*}_{p}$ has the negative contribution to symmetry energy and the effect is more pronounced with increasing the nucleon momentum \cite{Fe11}. The negative symmetry energy reduces the neutron emission, but is favorable to the energetic proton production. The case of $m^{*}_{n}<m^{*}_{p}$ just gives an opposite contribution. The influence of the isospin splitting of effective mass on the fast nucleon emission in heavy-ion collisions has also been investigated with the stochastic mean-field (SMF) model \cite{Ri05} and the improved quantum molecular dynamics (ImQMD) model \cite{Zh14}.

%%%%%%%%%%%%%%%%%%%%%%%%%%%%%%%%%%%%%%% figure 2 %%%%%%%%%%%%%%%%%%%%%%%
\begin{figure}
\begin{center}
{\includegraphics*[width=1.\textwidth]{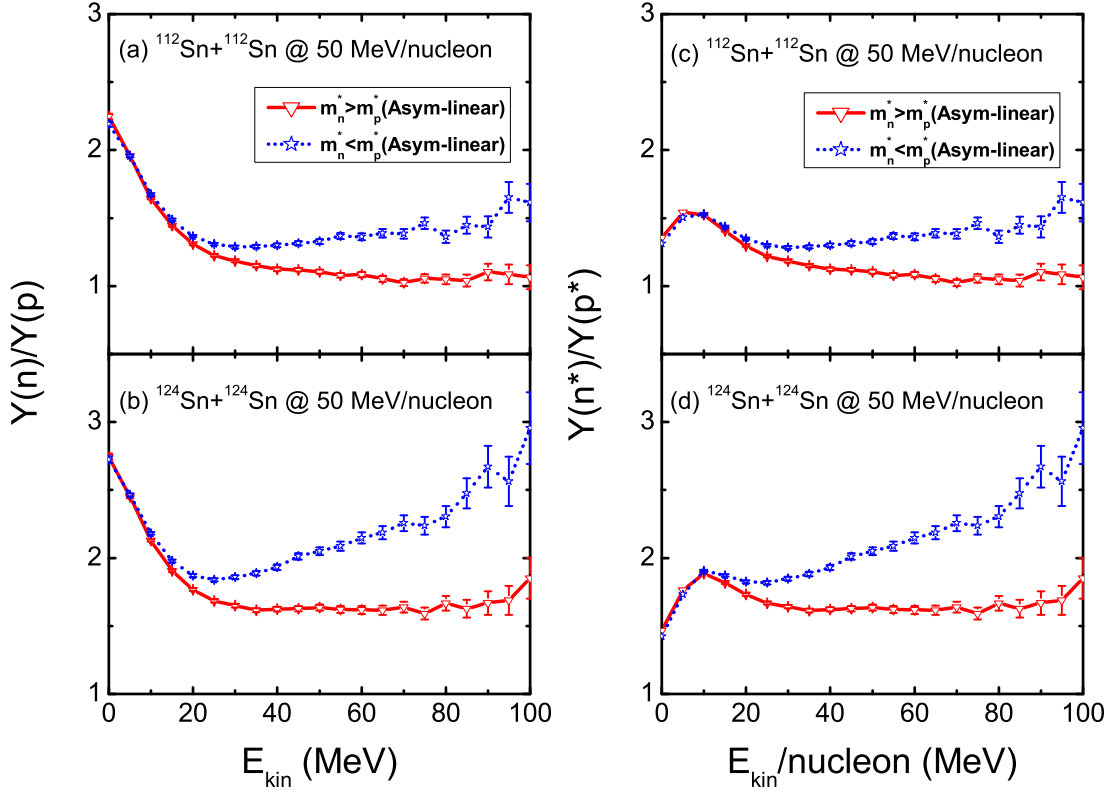}}
\end{center}
\caption{ Kinetic-energy spectra of neutron to proton ratios from the yields of free nucleons [panels (a) and (b)] and gas-phase fragments (nucleons, hydrogen and helium isotopes) [panels (c) and (d)] in the $^{112}$Sn+$^{112}$Sn and $^{124}$Sn+$^{124}$Sn reactions at 50 MeV/nucleon with the effective mass splittings of $m^{*}_{n}>m^{*}_{p}$ and $m^{*}_{n}<m^{*}_{p}$, respectively. }
\end{figure}
%%%%%%%%%%%%%%%%%%%%%%%%%%%%%%%%%%%%%%%%%%%%%%%%%%%%%%%%%%%%%%%%%%%%

%%%%%%%%%%%%%%%%%%%%%%%%%%%%%%%%%%%%%%% figure 3 %%%%%%%%%%%%%%%%%%%%%%%%
\begin{figure}
\begin{center}
{\includegraphics*[width=1.\textwidth]{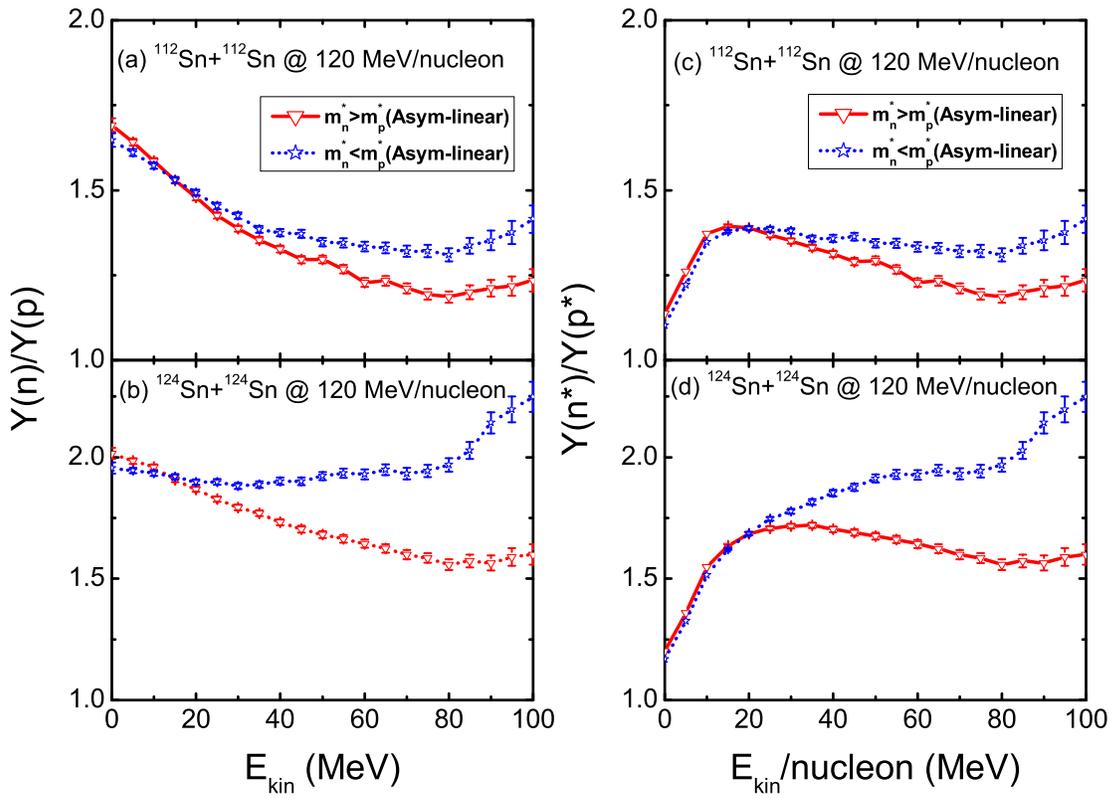}}
\end{center}
\caption{The same as in Fig. 2, but for the energy of 120 MeV/nucleon.}
\end{figure}
%%%%%%%%%%%%%%%%%%%%%%%%%%%%%%%%%%%%%%%%%%%%%%%%%%%%%%%%%%%%%%%%%%%%

%%%%%%%%%%%%%%%%%%%%%%%%%%%%%%%%%%%%%%% figure 4 %%%%%%%%%%%%%%%%%%%%%%%%
\begin{figure}
\begin{center}
{\includegraphics*[width=1.\textwidth]{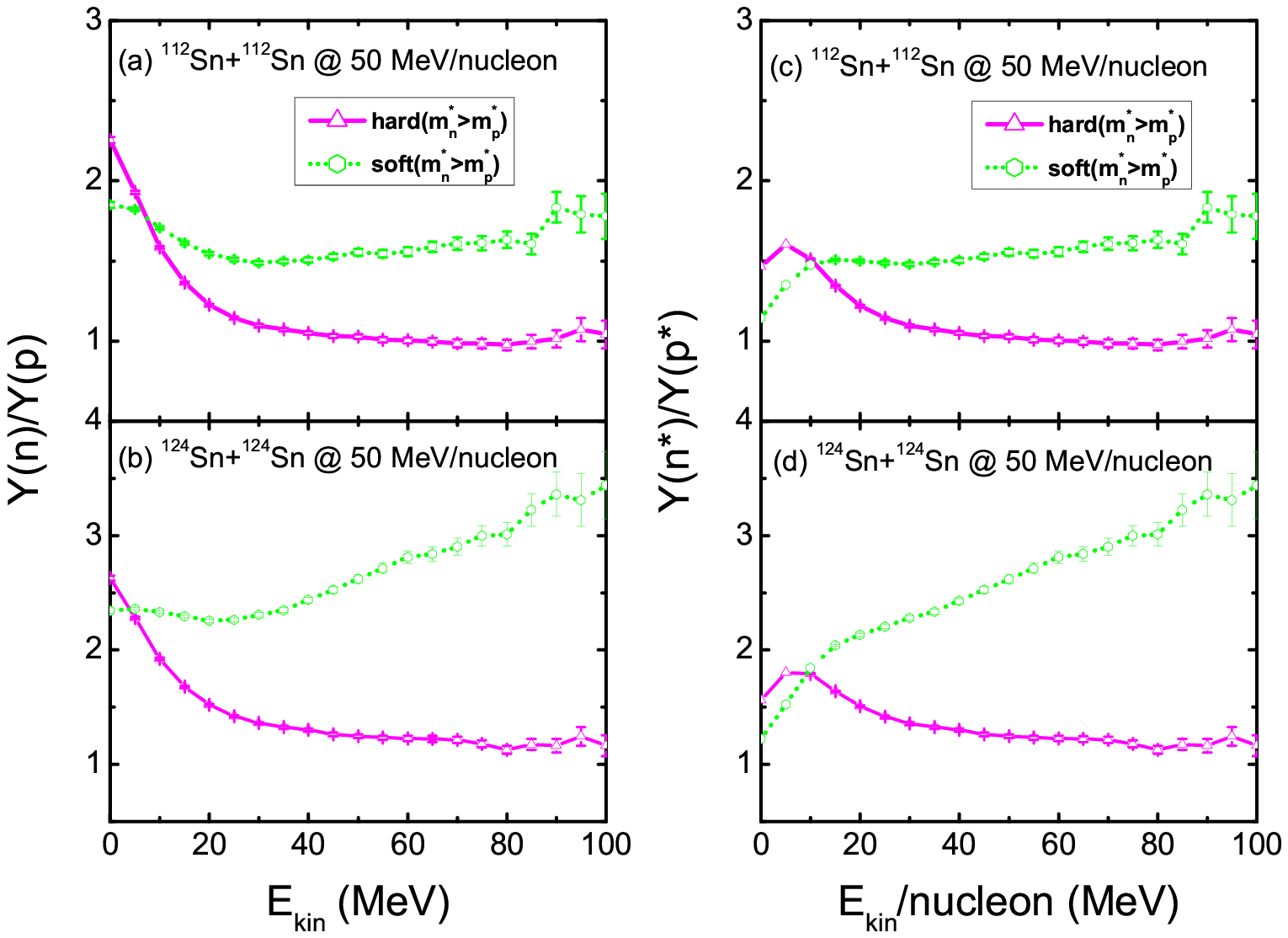}}
\end{center}
\caption{(Color online) The n/p ratios of free nucleons [panels (a) and (b)] and gas-phase fragments [panels (c) and (d)] as a function of kinetic energy with hard and soft symmetry energies, respectively.}
\end{figure}
%%%%%%%%%%%%%%%%%%%%%%%%%%%%%%%%%%%%%%%%%%%%%%%%%%%%%%%%%%%%%%%%%%%%

Both the isospin splitting of nucleon effective mass and symmetry energy impact the isospin dynamics in heavy-ion collisions. But the specific structure of isospin observable is different with the two quantities. Shown in Fig. 2 and in Fig. 3 is the neutron to proton (n/p) ratios from the free nucleons and gas-phase fragments in collisions of $^{112}$Sn+$^{112}$Sn and $^{124}$Sn+$^{124}$Sn at the incident energies of 50 MeV/nucleon and 120 MeV/nucleon, respectively. The same constraint condition as in Fig. 1 is used to analyze the transverse emission particles. It is found that the isospin splitting appears at the kinetic energies above 20 MeV/nucleon. The case of $m^{*}_{n}<m^{*}_{p}$ has a larger n/p ratio at the both incident energies. The effect is more pronounced for the neutron-rich system. The bump structure around the energy of 10 MeV/nucleon comes from the competition of free nucleons and light fragments to the n/p ratios. The light fragments with Z$\leq$2 are mainly produced at the low kinetic energy and have the smaller n/p ratios in comparison to the free nucleons \cite{Fe16}. At the kinetic energies above 30 MeV/nucleon, the yields of protons and neutrons are mainly contributed from the free nucleons. Impact of the stiffness of symmetry energy on the kinetic energy spectra is investigated in collisions of $^{112}$Sn+$^{112}$Sn and $^{124}$Sn+$^{124}$Sn at the incident energy of 50 MeV/nucleon as shown in Fig. 4. The difference of the hard ($\gamma_{s}$=2) and soft ($\gamma_{s}$=0.5) symmetry energy is obvious in the whole energy range. A flat structure in the $^{112}$Sn+$^{112}$Sn reaction appears with the soft case. The n/p ratios increase with the kinetic energy for the neutron-rich system.

%%%%%%%%%%%%%%%%%%%%%%%%%%%%%%%%%%%%%%% figure 5 %%%%%%%%%%%%%%%%%%%%%%%%
\begin{figure}
\begin{center}
{\includegraphics*[width=1.\textwidth]{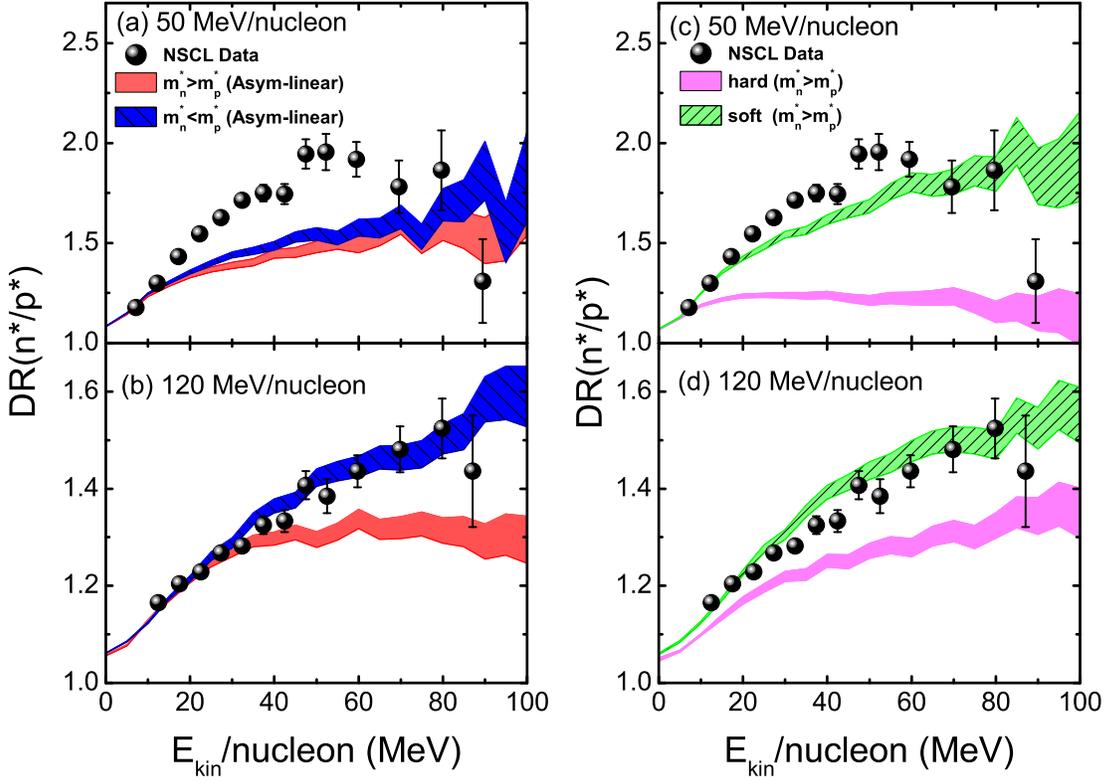}}
\end{center}
\caption{ The double neutron to proton ratios of gas-phase particles in collisions of $^{124}$Sn+$^{124}$Sn over $^{112}$Sn+$^{112}$Sn at the incident energy of 50 MeV/nucleon (upper panels) and 120 MeV/nucleon (lower panels) with different isospin splitting of effective mass and different stiffness of symmetry energy. The available data were measured at NSCL \cite{Co16}.}
\end{figure}
%%%%%%%%%%%%%%%%%%%%%%%%%%%%%%%%%%%%%%%%%%%%%%%%%%%%%%%%%%%%%%%%%%%%

%%%%%%%%%%%%%%%%%%%%%%%%%%%%%%%%%%%%%%% figure 6 %%%%%%%%%%%%%%%%%%%%%%%%
\begin{figure}
\begin{center}
{\includegraphics*[width=1.\textwidth]{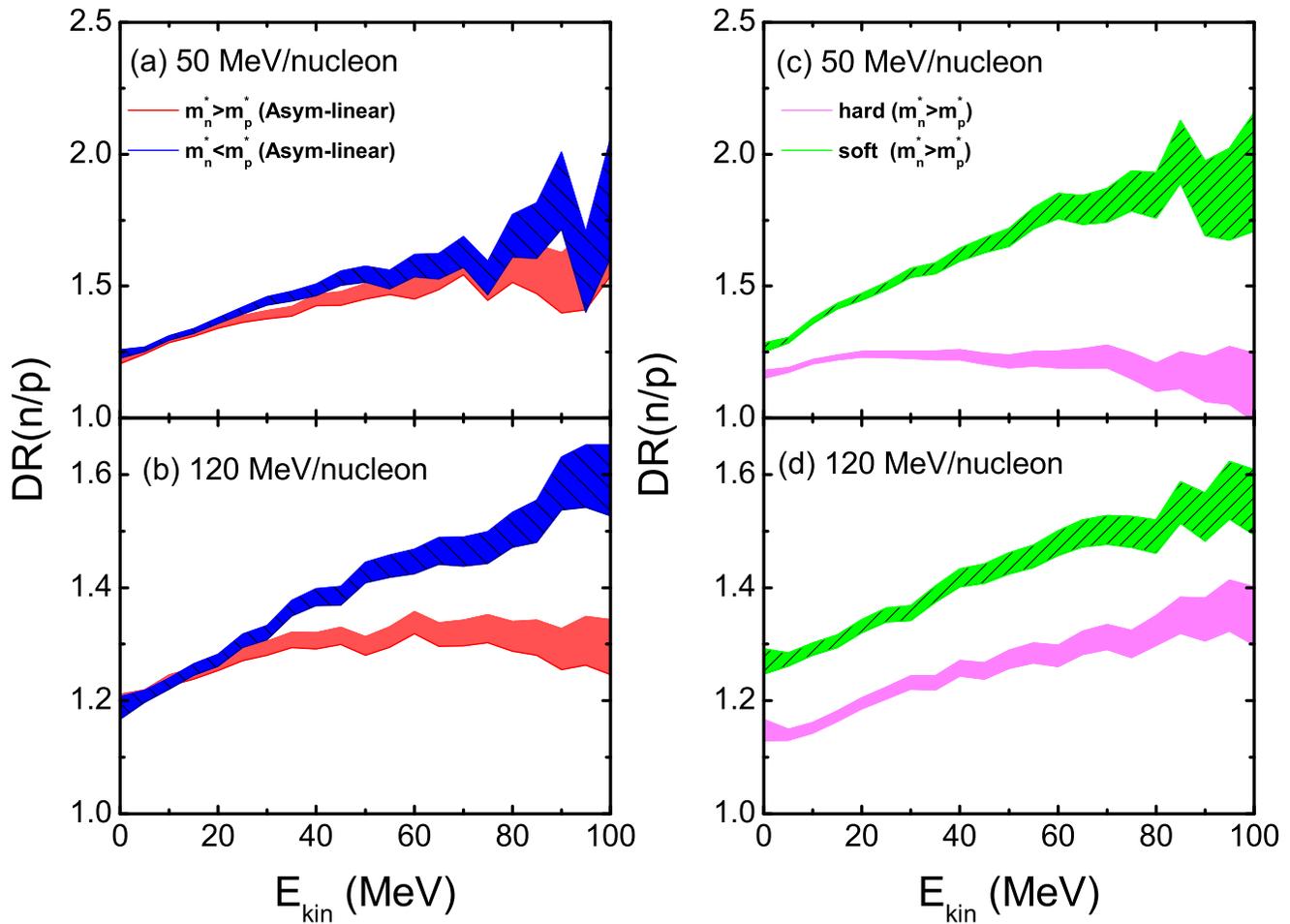}}
\end{center}
\caption{(Color online) The same as in Fig. 5, but for the double ratio spectra of free nucleons.}
\end{figure}
%%%%%%%%%%%%%%%%%%%%%%%%%%%%%%%%%%%%%%%%%%%%%%%%%%%%%%%%%%%%%%%%%%%%

Besides the effective mass and the stiffness of symmetry energy, the n/p ratios are also influenced by the Coulomb potential and the detector efficiencies of protons and neutrons in experiments. The double ratios of two isotopic systems would be nice probes for constraining the isospin splitting of nucleon effective mass and the symmetry energy beyond the saturation density from the experimental data. Shown in Fig. 5 is the double ratio spectra in collisions of $^{124}$Sn+$^{124}$Sn over $^{112}$Sn+$^{112}$Sn. The influence of the isospin splitting with a linear symmetry energy (left panels) and the stiffness of symmetry energy with the mass splitting of $m^{*}_{n}>m^{*}_{p}$ (right panels) on the spectra is compared with the new data at NSCL \cite{Co16}. It should be noticed that the effective mass splitting of neutron and proton in nuclear medium is pronounced the kinetic energies above 30 MeV/nucleon and the splitting of $m^{*}_{n}<m^{*}_{p}$ is nicely consistent with the available data at the beam energy of 120 MeV/nucleon. However, the difference of the both splittings on the spectra is very small at the beam energy of 50 MeV/nucleon. The symmetry energy effect is obvious and appears at the kinetic energy of 10 MeV/nucleon. A soft symmetry energy ($\gamma_{s}$=0.5) is constrained from the both incident energies. Furthermore, the difference of the soft and hard cases is more pronounced at the beam energy of 50 MeV/nucleon. The double ratio distributions from the free nucleons are also investigated as shown in Fig. 6. More pronounced effect from the stiffness of symmetry energy is observed, in particular at the kinetic energies below 30 MeV. Actually, the gas-phase particles at the high kinetic energies are mainly contributed from the free nucleons.

In summary, within the LQMD transport model we have investigated the isospin dynamics in heavy-ion collisions. The mass splitting of neutron and proton in nuclear medium and the stiffness of symmetry energy are constrained from the MSU data. The isospin splitting of nucleon effective mass is more pronounced at the beam energy of 120 MeV/nucleon. However, the difference of the soft and hard symmetry energies is obvious at the lower energy (50 MeV/nucleon). The soft symmetry energy with $\gamma_{s}$=0.5 and the splitting of $m^{*}_{n}<m^{*}_{p}$ are constrained from the double ratio spectra. The single n/p ratio decreases with the kinetic energy and then goes up at the case of $m^{*}_{n}<m^{*}_{p}$, in particular for the neutron-rich system. A flat structure appears with the soft symmetry energy until 30 MeV for free nucleons and the n/p ratio goes up monotonically with increasing the kinetic energy in the $^{124}$Sn+$^{124}$Sn reaction at the incident energies of 50 MeV/nucleon.

Fruitful discussions with Betty Tsang and Yingxun Zhang are grateful. This work was supported by the Major State Basic Research Development Program in China (No. 2014CB845405 and 2015CB856903), the National Natural Science Foundation of China (Projects Nos 11675226, 11175218, and U1332207) and the Youth Innovation Promotion Association of Chinese Academy of Sciences.

\end{document}